\documentclass[twocolumn,aps,prb,showpacs,preprintnumbers,amsmath,amssymb]{revtex4}
\usepackage{graphicx}
\usepackage{bm}
\usepackage{gensymb}
\usepackage{color}
\begin{document}
\title{Proposed low-energy model Hamiltonian for the spin-gapped system 
CuTe$_2$O$_5$}
\author{Hena Das$^{1}$, Tanusri Saha-Dasgupta$^{1}$, Claudius Gros$^{2}$ and Roser Valent\'\i$^{2}$}
\affiliation{$1$ Department of Material Sciences, S. N. Bose National Center for
 Basic Sciences, JD-III, Salt Lake City, Kolkata 700 098, India\\
$2$ Institut f\"{u}r Theoretische Physik, Universit\"{a}t Frankfurt, D-60438 Frankfurt, Germany}
\date{\today}
\begin{abstract}
Using first-principles electronic structure calculations based on the N$^{th}$ 
order muffin tin orbital (NMTO)-downfolding technique,  we derived the 
low-energy spin model for CuTe$_2$O$_5$. Our study 
reveals that this compound is a  2D coupled spin-dimer system with the 
strongest Cu-Cu interaction mediated by two O-Te-O bridges. We 
checked the goodness of our model by computing the magnetic 
susceptibility with the  
Quantum Monte Carlo technique and by comparing it with available 
experimental data. We also present magnetization and specific 
heat results which may be compared with future 
experimental investigations. Our derived model is in disagreement 
with a recently proposed model for this compound [J. Deisenhofer {\it et al}, 
Phys. Rev. B,{\bf 74} (2006) 174421]. The situation needs to be settled 
in terms of further experimental investigations.  
\end{abstract}

\maketitle
\section{\label{sec:introduction}Introduction }
Significant amount of effort both experimental and theoretical has 
been devoted in the last years to the investigation of the behavior of
low-dimensional quantum spin systems\cite{1}.
A crucial piece of information needed in the process of understanding
these systems is the connection between the underlying
chemical complexity
of the compound and the spin lattice.
Often, this relation is not obvious from structural
considerations and one needs to rely on {\it ab initio} based calculations
as we will show in the present work.

 Recently, in an attempt to analyze the effect of lone-pair cations
like Se$^{4+}$ or Te$^{4+}$ on the magnetic dimensionality of Cu$^{2+}$-based
systems,
the magnetic properties of CuTe$_2$O$_5$ were investigated\cite{2}. 
CuTe$_2$O$_5$ is structurally a Cu(II)-dimer system separated by Te ions. Magnetic 
susceptibility measurements show a maximum at T$_{max}$= 56.5\,K 
with an exponential drop at lower temperatures signaling 
the opening of a spin gap. The behavior at high temperatures follows the Curie
law with a    Curie-Weiss temperature 
of $\theta$=-41\,K\cite{1}, what indicates that the dominant interactions in this system
are antiferromagnetic.
 Electron spin resonance (ESR) observations suggest though that the structural 
dimers of CuTe$_2$O$_5$ do not coincide with the magnetic dimers\cite{2}. 
Fitting  the susceptibility data to different models, such as a pure 
dimer model, the alternating spin-chain model and the modified 
Bleaney-Bowers model, show equally good agreement of the experimental data\cite{2}.  
A detailed investigation of the magnetic exchange paths using the
 extended Huckel 
tight binding (EHTB) method  suggests that (i)
the strongest 
interaction is  between Cu ions which
are 6$^{th}$ nearest neighbors (J$_6$) and
is  of antiferromagnetic super-superexchange (SSE)
 type mediated by a O-Te-O bridge  and (ii)
that the next strongest interaction is of 
antiferromagnetic superexchange (SE) type within the structural dimer 
(J$_1$), yielding a ratio J$_1$/J$_6$=0.59\cite{2}. Based on 
these findings, the authors of Ref.\,\onlinecite{2} proposed an alternating spin chain model as the 
simplest possible model for CuTe$_2$O$_5$. 

Given the existence of many possible Cu-Cu  interaction paths in this
material whose relative importance may not be necessarily captured in
 EHTB study, we performed {\it ab initio} density
functional theory (DFT)
calculations and applied the NMTO-downfolding technique\cite{5}.  This technique
 has proven to be very successful  in deriving the underlying spin model of a
 large
number of 
low-dimensional quantum spin systems
including cuprates\cite{cuprates1,cuprates2},  vanadates\cite{vanadates} 
and titanates\cite{titanates}.
 Our calculations reveal that the strongest  Cu-Cu interaction in
 CuTe$_2$O$_5$ is
the one between 4$^{th}$ nearest neighbor mediated by 
two O-Te-O bridges (J$_4$) followed by the interaction
  mediated by a single O-Te-O bridge (J$_6$). This is in contrast with the
fact that J$_6$   was found in
  Ref.\,\onlinecite{2} 
 to be  the strongest interaction.  We also obtain that the Cu-Cu interaction within 
the structural dimer unit (J$_1$) is  rather weak as opposed 
to the findings of the EHTB study.  The underlying spin model for
CuTe$_2$O$_5$
 derived out of our calculations 
is therefore different from that suggested in 
Ref.\,\onlinecite{2}. We have also computed the
 magnetic susceptibility  for
the proposed model by performing Quantum Monte Carlo (QMC) simulations
 (stochastic 
series expansion\cite{qmc1,qmc2,qmc3}). Our  
results  show good agreement with 
the experimental observations. In view of the fact that the magnetic 
susceptibility  is often found to be an insensitive quantity to the details 
of the magnetic structure, we have also calculated temperature
and magnetic field dependent  magnetization as well as  
 the specific heat as a function of temperature. These results  need to be tested in terms of further 
experimental investigations to resolve the underlying microscopic model for
CuTe$_2$O$_5$ completely.

The paper is organized as follows: in section II and III we present
respectively  the crystal
structure and the {\it ab initio} DFT electronic structure
of CuTe$_2$O$_5$. In section IV we discuss the effective model
Hamiltonian obtained with the NMTO downfolding method. QMC results for
magnetic susceptibility, magnetization and specific heat are described in
section V and finally in section VI we present our conclusions.
%\vspace{2 in}
\section{\label{sec:crystal structure}Crystal structure }

\begin{figure}
\centering
\includegraphics[width=9cm]{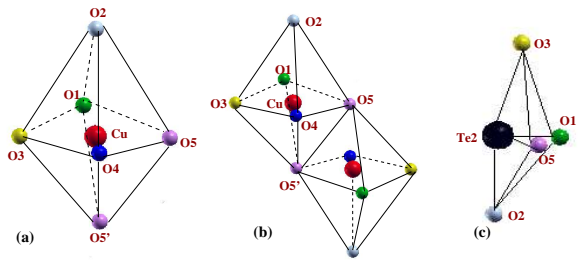}
\caption{\label{fig-1}(Color online) Building units of CuTe$_2$O$_5$. (a) CuO$_6$-distorted 
octahedron. (b) Cu$_2$O$_{10}$-structural dimer unit. (c) TeO$_4$-tetrahedra.}
\end{figure}

CuTe$_2$O$_5$ crystallizes in the monoclinic space group P21/c\cite{3} 
with lattice parameters $a$=6.871 \r{A}, $b$=9.322 \r{A}, $c$=7.602 \r{A}, 
and $\beta$=109.08\degree. It is built out of CuO$_6$ distorted  
octahedra (Fig.~\ref{fig-1}a), with six inequivalent oxygen
O1, O2, O3, O4, O5, O5$^{\prime}$ surrounding each Cu(II) ion. 
Each CuO$_6$ octahedron is elongated along the O2-O5$^{\prime}$ axis, 
with distances d$_{Cu-O5^{\prime}}$=2.303 \r{A} and d$_{Cu-O2}$=2.779 \r{A}. 
The Cu-O distances in the CuO$_4$ plane range from 
d$_{Cu-O5}$=1.948 \r{A} to d$_{Cu-O3}$=1.969 \r{A}. Two neighboring 
CuO$_6$ octahedra  share an edge to form a Cu$_2$O$_{10}$ structural 
dimer (Fig.~\ref{fig-1}b).   The oxygen octahedra of two Cu(II) ions 
within a given structural dimer are rotated by 180\degree with respect to each other.

\begin{figure}
\centering
\includegraphics[width=8cm]{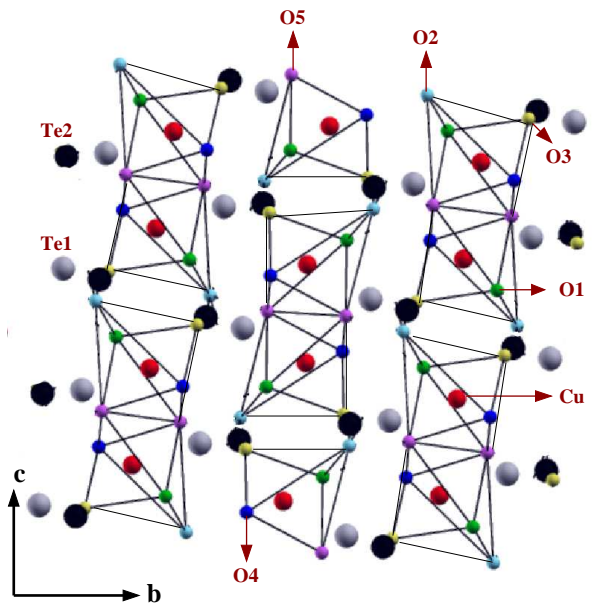}
\caption{\label{fig-2}(Color online) Crystal structure of CuTe$_2$O$_5$.
 The largest balls represent Te1 and Te2. Te1 and Te2 are shown in grey and 
black colors respectively. Cu atoms are represented by medium sized balls, 
situated at the centre of the distorted octahedra.  The smallest balls
denote the oxygen atoms.}
\end{figure}

The structural dimers form a chain-like structure running almost parallel to the
crystallographic {\it c} axis. These chains pile along the 
 crystallographic {\it b} axis (Fig.~\ref{fig-2}). The Te1 atoms
are situated between two successive Cu(II)-structural dimer chains, while the 
Te2 atoms are located in between two Cu$_2$O$_{10}$ structural dimers along 
a  given chain.
 The local oxygen environment of the Te atoms form a TeO$_4$ 
tetrahedra (Fig.~\ref{fig-1}c). The layers containing these chains in 
the  {\it bc} plane are stacked approximately along the  crystallographic 
{\it a} axis with hardly any connection between the layers. 

\section{\label{sec:band structure} electronic  structure }

\begin{figure}
\centering
\includegraphics[width=9cm]{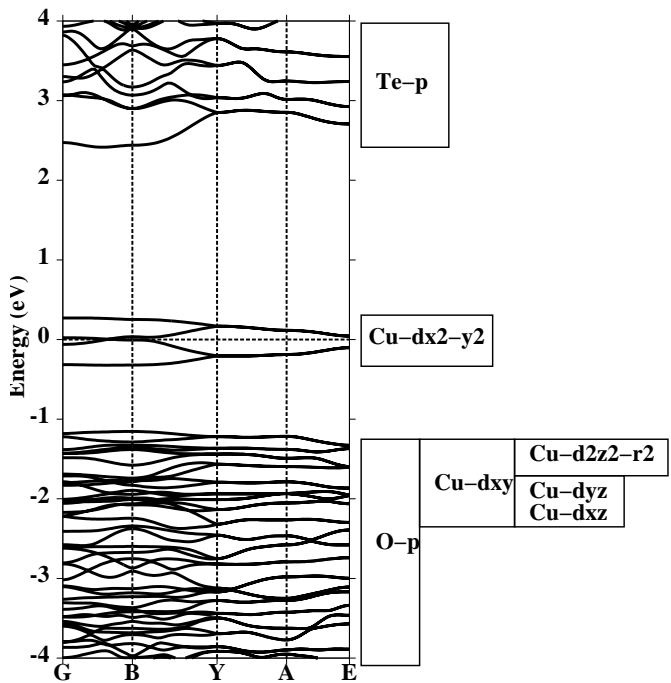}
\caption{\label{fig-3}LDA band structure of CuTe$_2$O$_5$ plotted along 
various symmetry directions of the monoclinic lattice. The zero of the energy 
has been set up at the LDA Fermi energy. The dominant orbital contributions 
in various energy ranges are shown in boxes on the right-hand side. The 
various Cu-d characters are shown in the local reference frame as
described in the text. }
\end{figure}

\begin{figure}
\centering
\includegraphics[width=8.3cm]{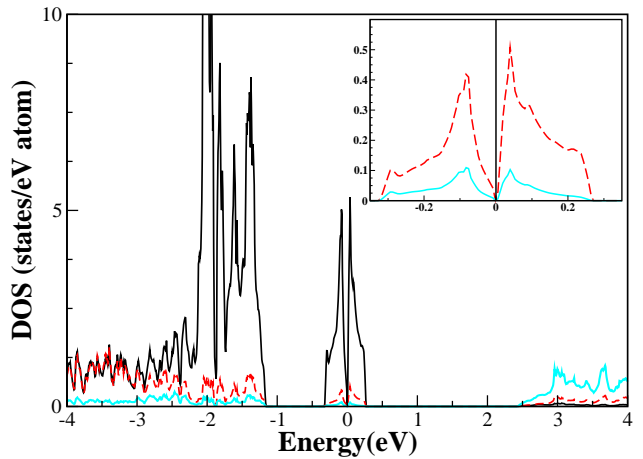}
\caption{\label{fig-4} (Color online) Partial density of states of Cu-$d$ (in
  black full lines), O-$p$ (in red dashed lines) and Te-$p$ (in cyan or gray
  full lines) orbitals, for CuTe$_2$O$_5$. The
inset shows the density of states for O-$p$ and 
Te-$p$ in the energy range close to E$_f$, dominated by Cu-$d_{x^2-y^2}$ character.}
\end{figure}

In order to analyze the electronic behavior of CuTe$_2$O$_5$ we  
carried out DFT calculations within the local density approximation by 
employing both the Wien2k code based on the full-potential linearized 
augmented plane wave (LAPW) method\cite{lapw} and the Stuttgart TBLMTO-47 code 
based on the linear muffin-tin orbital (LMTO) method \cite{4}. 
The calculated bandstructures agree well with each other within the allowed 
error bars of the various approximations involved in these two methods.  
Fig.~\ref{fig-3} and Fig.~\ref{fig-4} show the non-spin polarized band 
structures and density of states (DOS) respectively of CuTe$_2$O$_5$.
 The orbital contributions to the valence and conduction 
bands in the band structure and the DOS were determined by defining
 the 
local reference frame with the local z-axis pointing along Cu-O2 bond and the
local y-axis 
pointing almost parallel to the Cu-O5 bond.

The predominant feature of the band structure is the isolated manifold of four 
bands crossing the Fermi level (E$_f$), formed by Cu-$d_{x^2-y^2}$ orbitals
corresponding to the four Cu atoms 
 in the unit cell, admixed with O-$p$ states.
These bands are half filled and separated from the low lying O-$p$ and 
non-$d_{x^2-y^2}$  Cu 
 valence bands by a gap of about 0.8 eV and from the Te-$p$-dominated
 high lying conduction bands by a gap of about 2.2 eV, with the
zero of energy set at the LDA Fermi level. We note that in the low energy 
scale, the LDA results lead to  a metallic state. Introduction of missing 
correlation effects beyond LDA, for instance with the LDA+U
functional,  is expected to drive the system insulating, as our LDA+U calculations
corroborated.

 The valence band shows 
 Cu $d_{xy}$, $d_{yz}$, $d_{zx}$ and $d_{3z^2-r^2}$ character
dominated  bands
 in the energy range between -2.2 eV and -1.2 eV  while the O-$p$-dominated 
bands are in the energy range between -4 eV and -1.2 eV. 
The contribution of O2 character in the conduction bands crossing the 
Fermi level is found to be small compared to other oxygen because 
of the large Cu-O2 bond length. The Te1-$p$ and Te2-$p$ states 
 show a non-negligible contribution to the bands crossing the 
Fermi energy, as pictured in the inset of Fig.~\ref{fig-4} and play an
important role 
in mediating the Cu-Cu exchange interaction as will be demonstrated  in what 
follows.  

\section{\label{sec:low energy hamiltonian}Low energy Hamiltonian - Effective Model }

A powerful technique to construct a low-energy, tight binding (TB) Hamiltonian 
starting from a LDA band structure, is given by the NMTO downfolding method 
\cite{5}. This method derives a low-energy Hamiltonian by an energy selective, 
downfolding process that integrates out the high energy degrees of freedom. 
The  low energy Hamiltonian is then defined in the basis of effective orbitals 
constructed via the integration out process. This process takes into account the 
proper renormalization effect from the orbitals that are being downfolded. The 
accuracy of such process can be tuned by the choice of the number of energy 
points (N), used for the selection of downfolded bands. If the low-energy bands 
form an isolated set of bands, as is the case under discussion, the constructed 
effective orbitals, the NMTOs, span the Hilbert space of Wannier functions or 
in other words, the effective orbitals are the Wannier functions corresponding 
to the low-energy bands. The real space representation of the downfolded 
Hamiltonian H=$\sum t_{ij} (c_i^\dagger c_j + h.c) $ in the Wannier function 
basis gives the various hopping integrals $t_{ij}$ between the effective 
orbitals.

\begin{figure}
\centering
\includegraphics[width=6.3cm]{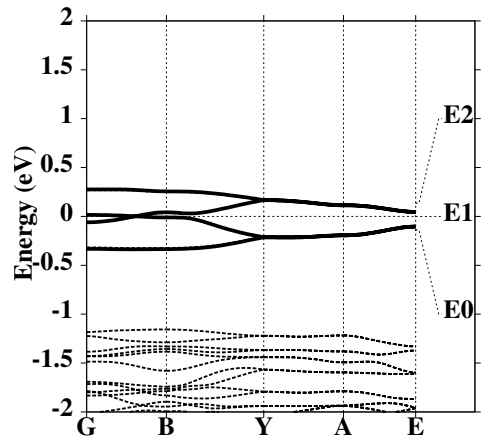}
\caption{\label{fig-5}Bands obtained with downfolded Cu-$d_{x^2-y^2}$ basis (solid lines) compared to full LDA band structure (dashed lines). E$_0$, E$_1$ and E$_2$ mark the energy points used in NMTO calculation.}
\end{figure}

For the present compound we have derived the low energy Hamiltonian defined 
in the basis of the effective Cu-$d_{x^2-y^2}$ orbitals, by keeping only the 
$d_{x^2-y^2}$ orbital for each Cu atom in the unit cell and integrating out 
all the rest. We show the downfolded bands in Fig.~\ref{fig-5} in comparison to 
the full LDA band structure. With the choice of three energy points E$_0$, 
E$_1$ and E$_2$, downfolded bands are indistinguishable from the 
Cu-$d_{x^2-y^2}$ dominated bands of the full LDA calculation.

\begin{figure}
\centering
\includegraphics[width=6.2cm]{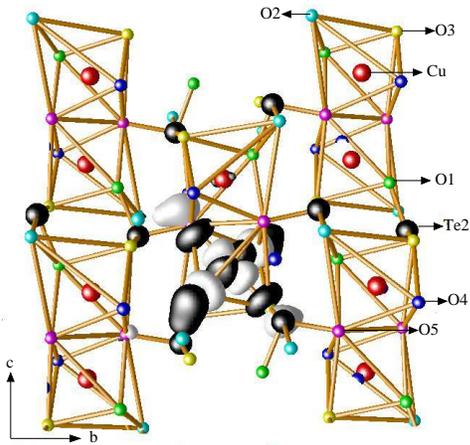}
\caption{\label{fig-6} (Color online) Effective Cu$d_{x^2-y^2}$ orbital with lobes of opposite 
signs colored as black and white. The $d_{x^2-y^2}$ orbital is defined with 
the choice of local reference frame as described in the text.} 
\end{figure}

The corresponding Wannier function is plotted in Fig.~\ref{fig-6}. The central 
part has the 3$d_{x^2-y^2}$ symmetry with the choice of the local coordinate 
system as stated above, while the tails are shaped according to O-$p_x$/$p_y$.
The Cu-$d_{x^2-y^2}$ orbital forms strong pd$\sigma$ antibonds with the 
O-$p_x$/$p_y$ tails. O-$p_x$/$p_y$ tails bend towards the Te2 atom,
what  
indicates the importance of hybridization effects from the Te cations and 
enhances the Cu-Cu interaction placed at different structural dimers 
Cu$_2$O$_{10}$.  

\begin{figure}
\centering
\includegraphics[width=8.8cm]{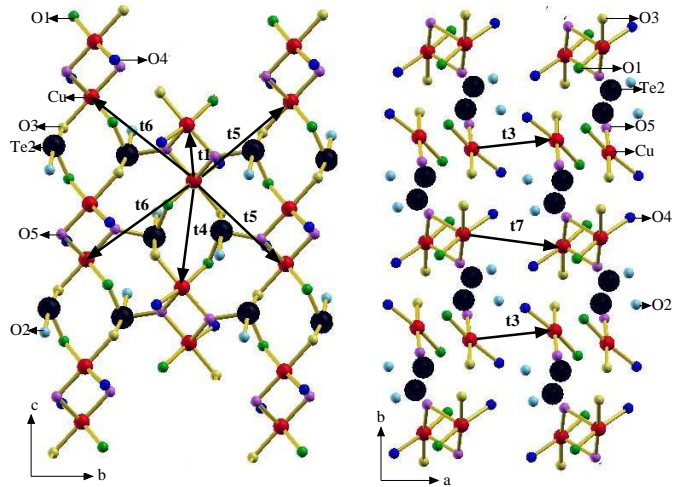}
\caption{\label{fig-7}(Color online)  Cu-Cu interaction paths $t_n$.  
The color convention is the same as Fig.~\ref{fig-2}.}
\end{figure}

Table-I shows the various dominant effective hopping integrals $t_{ij}$ 
 with magnitude greater than 1 meV between the Cu(II) ions at sites {\it i} and 
{\it j}. The notation for the various hoppings is shown in Fig.~\ref{fig-7}
where the subindex of t$_n$ corresponds to the $n^{th}$ Cu neighbors.
The 
strongest hopping integral, {\it t$_4$}, is found to be between those two 
Cu(II) ions which are placed at different structural dimers 
and the interaction is mediated by two O-Te-O bridges. 
{\it t$_1$}, which denotes the hopping integral  between two Cu(II) ions 
situated within the same structural dimer unit, is found to be about 1/3 of the 
strongest hopping integral ({\it t$_4$}). The second strongest  hopping term, 
{\it t$_6$}, mediated by one O-Te-O bridge is about 1/2 of  
{\it t$_4$}. Fig.~\ref{fig-7}b shows the interaction paths in the {\it ab} plane, 
which are weak in general and can be neglected.  In particular,  we 
mention as examples the hopping integrals {\it t$_3$} and {\it t$_7$}, which are approximately 
1/10 and 1/25 of the strongest hopping term ({\it t$_4$}) respectively. In the 
following we discuss the origin of the various dominant interaction paths.      

\begin{table}
\caption{Cu-Cu hopping parameters corresponding to the downfolded
  Cu-$d_{x^2-y^2}$ Hamiltonian
 in NMTO-Wannier function basis. Only hopping integrals of strength 
larger than 1meV are listed.}

\begin{tabular}{|c|c|c|c|c|c|}
\hline
 hopping   &   Cu-Cu distances  &    Hopping parameters  \\
           &       in \r{A}     &            in meV       \\
\hline 
{\it t$_1$} & 3.18 & 38.7 \\
\hline
{\it t$_3$} & 5.32 & 11.0  \\
\hline
{\it t$_4$} & 5.58 & 112.9  \\
\hline
{\it t$_5$} & 5.83 & 13.7   \\
\hline
{\it t$_6$} & 6.20 & 59.9   \\
\hline
{\it t$_7$} & 6.43 & 4.9    \\
\hline   
\end{tabular} 
\end{table}

\subsection{Strongest hopping term -{\it t$_4$}}

\begin{figure}
%\centering
\includegraphics[width=6.2cm]{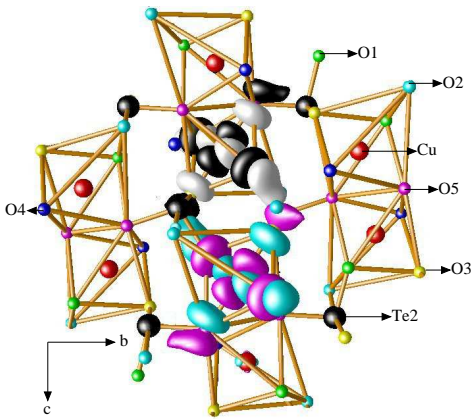}
\caption{\label{fig-8}(Color online) Effective orbital corresponding to the downfolded NMTOs,
placed at two Cu sites situated at two different structural dimer units 
corresponding to the {\it t$_4$} interaction. Lobes of orbitals placed at 
different Cu sites are colored differently. Lobe colored black (white) at one 
Cu site represents the same sign as that colored magenta (cyan) at other 
Cu site.}
\end{figure}

The strongest hopping term, {\it t${_4}$}, mediated by two O-Te-O bridges 
is associated to a  Cu-O-Te-O-Cu super-superexchange (SSE) path generating the 
spin-spin coupling J${_4}$. The strength of a SSE interaction
 through an 
exchange path of type Cu-O-L-O-Cu (e.g., L=Te) depends sensitively on how the 
O-L-O linkage orients the two magnetic orbitals 
(i.e the $d_{x^2-y^2}$ orbitals) centered at two Cu sites and also on how the 
tails of the magnetic orbitals, which have contributions of the orbitals of the ligand atom 
L, are oriented with respect to the central part. In Fig.~\ref{fig-8} 
we show the Wannier function plot corresponding to $t_4$, where the effective Cu-$d_{x^2-y^2}$-like 
Wannier orbitals are  at the Cu sites between which we have found the 
strongest interaction. The O-p$_x$/p$_y$ tails bend towards the Te atoms 
forming O-Te-O ligand paths which are   responsible for the strong Cu-Cu 
bonding.

\subsection{Second strongest hopping term -{\it t$_6$}} 
     
\begin{figure}
%\centering
\includegraphics[width=6.2cm]{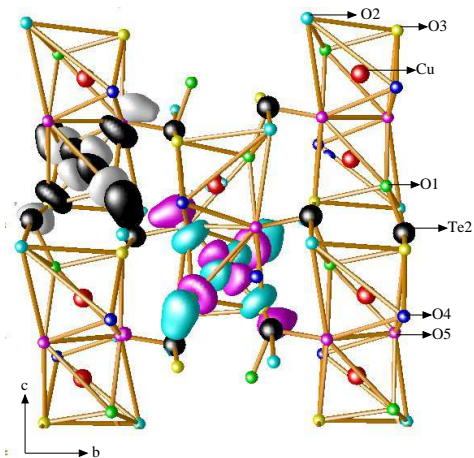}
\caption{\label{fig-9}(Color online) Effective orbital corresponding to the downfolded NMTOs,
placed at two Cu sites situated at two different structural dimer units 
corresponding to the t$_6$ hopping term. Color convention is same as in Fig.~\ref{fig-8}.}
\end{figure}

The hopping integral {\it t$_6$} describes  the next strong Cu-Cu 
interaction path, which is mediated via one O-Te-O bridge and responsible for 
the Cu-O-Te-O-Cu SSE interaction generating the spin-spin coupling J$_6$. 
Fig.~\ref{fig-9} shows the Wannier plots of the Cu-$d_{x^2-y^2}$ downfolded 
NMTOs.  Here the oxygen tails bend towards the interconnecting 
TeO$_2$ unit to provide an interaction pathway between the two Cu sites
as in the $t_4$ path. 
However the strength of this  interaction is expected to be weaker than 
{\it t$_4$} since there is only one, instead of  two O-Te-O interaction paths.

\subsection{Structural intradimer hopping term-{\it t$_1$}}

\begin{figure}
%\centering
\includegraphics[width=6.2cm]{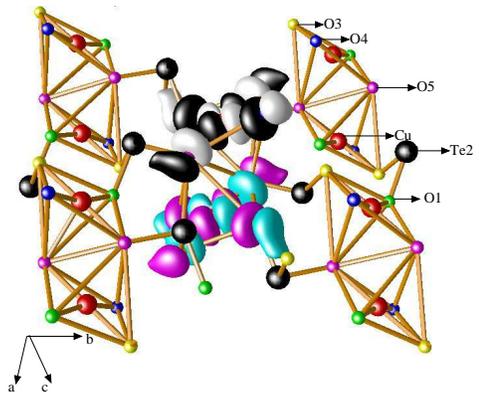}
\caption{\label{fig-10}(Color online) Cu-d$_{x^2-y^2}$ downfolded NMTOs, placed at two Cu 
sites situated within same structural dimer. The O2 sites with long Cu-O2 bond 
lengths have been removed 
for better view. Color convention is same as in 
Fig.~\ref{fig-8}.}
\end{figure}

{\it t$_1$} corresponds to the intradimer Cu-Cu interaction path which is mediated by 
O5-O5$^{\prime}$ atoms. In Fig.~\ref{fig-10} we show the Wannier function plot,
 where the effective Cu-$d_{x^2-y^2}$ like Wannier orbitals are placed at the 
Cu sites of the same structural dimer unit. As we stated above, each structural 
dimer unit is made of two edge sharing CuO$_6$ distorted octahedra. In the 
case of the first octahedron O5 is situated on the basal plane of the octahedron 
and O5-p$_x$/p$_y$ form pd$\sigma$ antibond with the Cu-d$_{x^2-y^2}$ orbital, 
whereas O5$^{\prime}$ is situated at the apical position for this octahedron.
 The reverse is true for the second octahedron. Therefore Cu-d$_{x^2-y^2}$ 
orbitals of two Cu$^{2+}$ sites placed at the same structural dimer unit are 
misaligned, what is responsible for the weak Cu-Cu intradimer interaction. As 
the Cu-O5-Cu and Cu-O5$^{\prime}$-Cu angles turn out to be 96.76 degree, 
the Cu-Cu interaction within the structural dimer, which is weak in general, 
is in the borderline where a sign change of exchange interaction from 
antiferromagnetic to ferromagnetic may occur\cite{fm-afm}.   

\section{\label{sec:susceptibility}Susceptibility, Magnetization and Specific heat}

\begin{figure}[h]
\includegraphics[width=7cm]{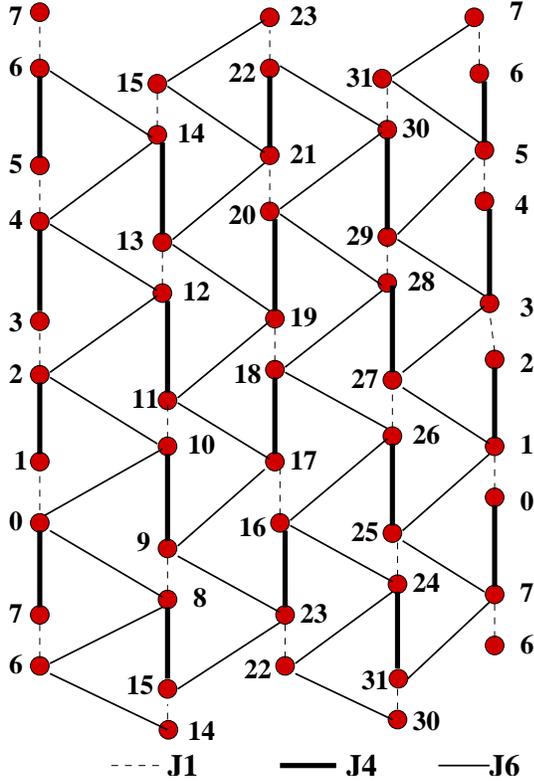}
\caption{\label{fig-11} The 2D coupled dimer model shown on a
8 x 4 (32 site) lattice. Periodic boundary conditions have been applied on
both directions. The thick, thin and dashed lines represent the
strongest J$_{4}$, the next strong J$_{6}$ and the weak structural
intra-dimer interaction J$_{1}$ respectively. The site index $k$ is given by
$k = N_1 * l + m$, where m runs over number of rows in the
 square lattice ($m = 0, 1,   \ldots N_1$),
and l runs over the number of columns in the square lattice ($l = 0, 1,   \ldots
N_2$) .}
\end{figure}

The description of the spin model
 for CuTe$_2$O$_5$ as obtained from the 
NMTO-downfolding technique turns out to be that of a system of coupled dimers in a two dimensional (2D) grid 
 (see Fig.~\ref{fig-11}). In order to check the goodness of our proposed 
model, we have calculated the magnetic susceptibility as well as magnetization 
and specific heat properties  by considering the following
spin-1/2 Heisenberg 
model on a $N_1 \times N_2$ lattice:

\begin{eqnarray}
H  & = &  J_1 \displaystyle \sum_{j=0}^{N_2-1}
 \sum_{i=0}^{\frac{N_1}{2} -1}
 ({\bf S}_{2i,j}{\bf S}_{2i+1,j}) +
 J_4 \displaystyle \sum_{j=0}^{N_2-1}
 \sum_{i=0}^{\frac{N_1}{2} -1}
 ({\bf S}_{2i+1,j}{\bf S}_{2i+2,j}) \nonumber \\
 & + & J_6 \displaystyle \sum_{j=0}^{\frac{N_2}{2}-1} \sum_{i=0}^{\frac{N_1}{2} -1} [ 
({\bf S}_{2i,2j}{\bf S}_{2i,2j+1}+
{\bf S}_{2i,2j}{\bf S}_{2i+2,2j+1})\nonumber \\
 & + & ({\bf S}_{2i+1,2j}{\bf S}_{2i+1,2j-1}+{\bf S}_{2i+1,2j}{\bf S}_{2i+3,2j-1}) ]
\label{model}
\end{eqnarray}

where J$_1$, J$_4$ and J$_6$ are the exchange integrals corresponding to the hopping 
paths {\it t$_1$}, {\it t$_4$} and {\it t$_6$} respectively. Interestingly,
this model reduces to the model grid that describes the magnetic
behavior of CaCuGe$_2$O$_6$ when $J_1 = 0$.
In that case, the two dimensional model has (using the present notation),
 two critical points at
$J_6$$\approx$$-0.9 J_4$ and $J_6$$\approx$$0.55 J_4$\cite{cuprates1}.

The analysis of  model Eq.~\ref{model}  has been done by the Quantum Monte Carlo
method (stochastic series 
expansion\cite{qmc1,qmc2,qmc3}) on a 20 x 20 lattice. While the NMTO downfolding technique 
gives  us an estimation for hopping parameters, it does not provide 
directly values of exchange integrals. The exchange coupling, J, 
can be expressed in general as a sum of antiferromagnetic and ferromagnetic 
contributions J=J$^{AFM}$+J$^{FM}$. In the limit of large correlation, 
typically valid for Cu based system, the antiferromagnetic contribution, 
J$^{AFM}$, is related to the hopping integral {\it t} by the second order 
perturbation relation J$^{AFM} = 4t^2/U$, where U is the effective 
on-site Coulomb repulsion. In absence of a satisfactory approach for 
computing J directly, in the following we considered the NMTO 
downfolding inputs to built up the model and starting point for 
relative estimates of various exchange interactions. We define the parameters,

\begin{eqnarray} 
\alpha_1 = \frac{J_6}{J_4}, \quad
\alpha_2 = \frac{J_1}{J_4}
\label{ratio}
\end{eqnarray} 
which measure the ratio of the inter-dimer J$_6$  and structural intra-dimer
J$_1$ 
interactions with respect to the exchange interaction which was
suggested from the downfolding calculations to be the strongest J$_4$.

\begin{figure}
\includegraphics[width=7cm]{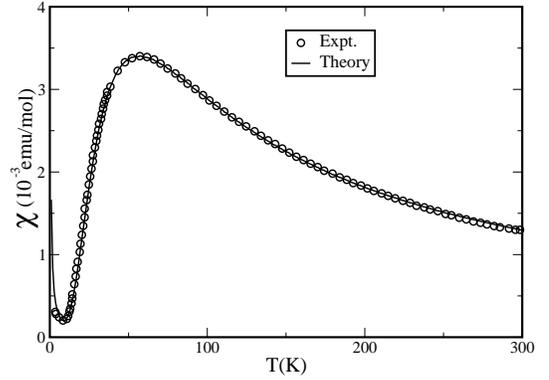}
\caption{\label{fig-12} Temperature dependence of magnetic susceptibility for 
CuTe$_2$O$_5$.  The circles corresponds to experimental data\cite{2} and 
the solid line correspond to calculated susceptibility based 
on a 2D-coupled dimer model.}
\end{figure}

The optimal values of 
$\alpha_1$ and  $\alpha_2$ as well as the strength of the primary interaction 
J$_4$ and the effective $g$ factor are obtained by fitting the QMC results for the 
susceptibility:

\begin{eqnarray}
\chi^{th}= \langle(S^z-\langle S^z\rangle)^2\rangle
\end{eqnarray}
 ($\mu_B$ and $k_B$ denote the Bohr magneton and the
Boltzmann constant respectively) 
 with the 
experimental susceptibility (in [emu/mol])
 at intermediate to high temperatures via\cite{kluemper} 
$\chi = 0.375 (g^2/J)\chi^{th}$. To simulate the low temperature region of 
the susceptibility data we include the respective Curie contribution 
from impurities as $\chi^{CW} = C_{imp}/T$. The calculated  susceptibility 
in comparison to experimental data is shown in
Fig.\ref{fig-12}. 

\begin{figure}
\includegraphics[width=7cm]{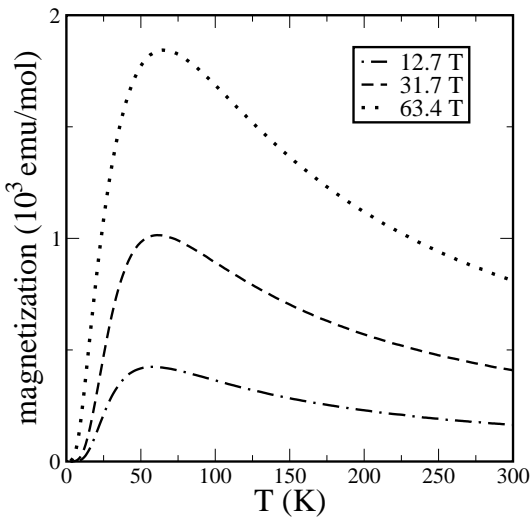}
\caption{\label{fig-13} Magnetization plotted as a function of temperature for the 2D-coupled 
dimer model of CuTe$_2$O$_5$ in an applied magnetic field of strengths h/J=0.2, 
0.5, 1.0 (bottom to top)  which correspond to $H= 12.7, 31.7, 63.4$ Teslas.
}
\end{figure}

\begin{figure}
\includegraphics[width=7cm]{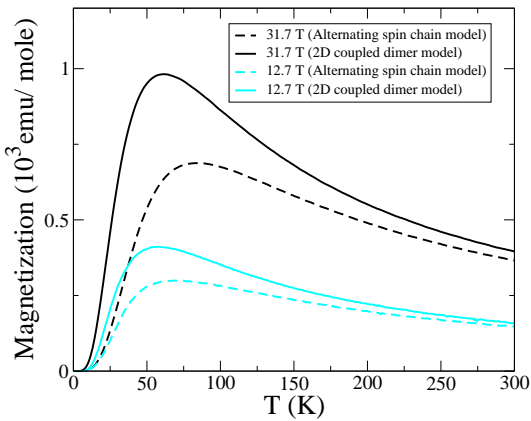}
\caption{\label{mag_comp} (Color online) Magnetization plotted as a function of temperature
  for the 2D-coupled
dimer model of CuTe$_2$O$_5$ and the alternating chain model
of Ref. \protect\onlinecite{2} for two values of the magnetic field
$H = 12.7$ Teslas
and $H =31.7$ Teslas. 
}
\end{figure}

The best fit corresponds to the intra-dimer exchange integral J$_4$=92.4 K, 
very close to the value proposed by J. Deisenhofer {\it et al}\cite{2}
for the strongest dimer coupling. The optimal 
value of the $g$ factor=2.17 was  found to be slightly larger than the spin only 
value of $g$=2,  in agreement with ESR measurement. The optimal values 
for the coupling ratios in Eq.~\ref{ratio} are  
 found to be $\alpha_1$ = 0.27 and $\alpha_2$ = 0.07, rather close to the 
estimates, 0.28 and 0.11 respectively, obtained using the second-order 
perturbation relationship between exchange interaction (J) and the hopping 
integral ({\it t}) given by the NMTO-downfolding study. The weak J$_1$ 
interaction turned out to be of antiferromagnetic nature giving rise to 
a positive sign for $\alpha_2$. 
With the stochastic series expansion  implementation of the quantum Monte Carlo 
method it is possible to simulate quantum spin models in an external field. 
In Fig. ~\ref{fig-13}, we present the  computed magnetization as a function
 of temperature $M(T)$ for various magnetic fields' strengths and  
in Fig.~\ref{mag_comp} we show the comparison of 
 $M(T)$  for the model proposed
in this work and the alternating chain model of Deisenhofer {\it et al.}\cite{2}
for $H=12.7$ Teslas and $H=31.7$ Teslas.
Two models show distinctly different behavior at moderate to high
magnetic fields.
 
\begin{figure}
\includegraphics[width=7cm]{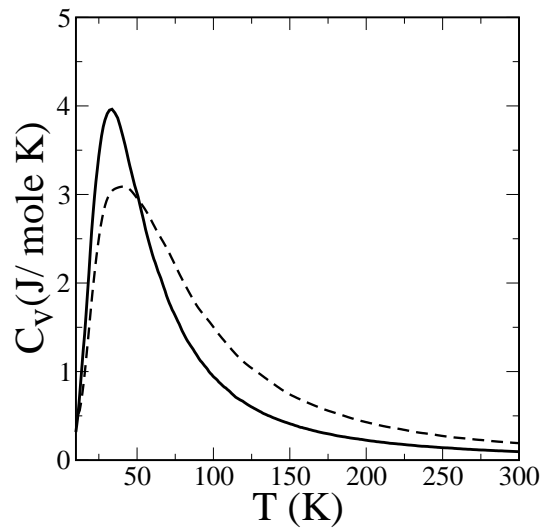}
\caption{\label{fig-14} Specific heat
 plotted as a function of 
temperature for CuTe$_2$O$_5$ for the 2D-coupled dimer model (solid line) and
the model of Ref. \protect\onlinecite{2} (dashed line)
.}
\end{figure}

 We also  calculated the specific heat $C_v(T)$ for
both models and the results are presented in Fig.~\ref{fig-14}.
While the overall qualitative shapes of the $C_v(T)$ versus temperature
curves for both models are similar,
 there
are important quantitative distinctions which capture the
different nature of the models. 

 Our above computed 
thermodynamic quantities  
provide a useful framework to test the validity of our 
proposed model in terms of further experimental measurements.
   
\section{\label{sec:conclusion}Conclusion }

The analysis of the electronic structure of CuTe$_2$O$_5$ by first-principles 
NMTO-downfolding calculations as well as the calculation and examination of 
susceptibility data by the QMC method leads to a unique description of  this system as a 2D 
coupled dimer model.
 The strongest Cu-Cu interaction is
 between Cu pairs belonging to different structural dimer 
units and connected by two O-Te-O bridges. Two additional in-plane 
interactions of about 1/3 and 1/10 of the strongest interaction have 
been found; the  latter one being the structural intra-dimer interaction. This is 
in disagreement with recent theoretical considerations
in Ref. \onlinecite{2}, 
which suggest the CuTe$_2$O$_5$ system as an alternating spin chain system 
with strong intra- and inter-dimer coupling. Based on our proposed model, 
we have also calculated the magnetization and specific heat which may be 
compared with new experimental measurements. We hope that our work 
will stimulate further experimental studies.     

\section{\label{sec:Aacknowledgment}Acknowledgment}
We would like to thank P. Lemmens and J. Deisenhofer for discussions and for 
bringing the problem into our attention. HD and TSD 
acknowledge MPG-India partner group program for the collaboration.
C.G. and R.V thank the German Science Foundation (DFG) for financial
support through the TRR/SFB 49 program.

%\section{\label{sec:references}References}

\end{document}